\documentclass[a4paper]{jpconf}
\usepackage{amsmath}  
\usepackage{graphicx,amssymb}
\usepackage{microtype}
\usepackage{xcolor}

\usepackage[square,comma,numbers,sort&compress]{natbib}

\makeatletter
\renewcommand\NAT@bibsetnum[1]{\settowidth\labelwidth{\@biblabel{#1}}%
   \setlength{\leftmargin}{\bibindent}%
   \addtolength{\leftmargin}{\dimexpr\labelwidth+\labelsep\relax}%
   \setlength{\itemindent}{-\bibindent}%
   \setlength{\listparindent}{\itemindent}
\setlength{\itemsep}{\bibsep}\setlength{\parsep}{\z@}%
   \ifNAT@openbib
     \addtolength{\leftmargin}{\bibindent}%
     \setlength{\itemindent}{-\bibindent}%
     \setlength{\listparindent}{\itemindent}%
     \setlength{\parsep}{0pt}%
   \fi
}
\makeatother

\usepackage[bookmarks]{hyperref}

\def\Im{\mathop{\rm Im}\nolimits}
\def\Re{\mathop{\rm Re}\nolimits}
\def\Tr{\mathop{\rm Tr}\nolimits}

\def\rmi{{\rm i}}

\def\full{\raise2pt\hbox{\vrule height.7pt depth0pt width1.5em}}
\def\shortdash{\raise2pt\hbox{%
\vrule height.7pt depth0pt width.15em\hskip.12em%
\vrule height.7pt depth0pt width.15em\hskip.12em%
\vrule height.7pt depth0pt width.15em\hskip.12em%
\vrule height.7pt depth0pt width.15em\hskip.12em%
\vrule height.7pt depth0pt width.15em%
}}
\def\thickfull{\raise2pt\hbox{\vrule height1.3pt depth0pt width1.5em}}
\def\dotted{{\tiny\raise1pt\hbox{$\bullet\bullet\bullet$}}}

\definecolor{dred}{rgb}{0.65,0.0,0.0}
\definecolor{dblue}{rgb}{0.14,0.27,0.62}

\begin{document}
\title{Theory of valence-band and core-level photoemission from
  plutonium dioxide}

\author{Jind\v rich Koloren\v c${}^1$, Agnieszka L. Kozub${}^{1,2}$ and
  Alexander B. Shick${}^1$}

\address{${}^1$~Institute of Physics, Academy of Sciences of the Czech
  Republic, Na Slovance 2, \phantom{${}^1$~}182$\,\,$21~Prague, Czech Republic}
\address{${}^2$~Faculty of Applied Physics and Mathematics, Gdansk
  University of Technology, \phantom{${}^2$~}Narutowicza~11/12, 80-233
  Gdansk, Poland}

\ead{kolorenc@fzu.cz}

\begin{abstract}
The correlated-band theory implemented as a combination of the
local-density approximation with the dynamical mean-field theory is
applied to PuO${}_2$. An insulating electronic structure, consistent
with the experimental valence-band photoemission spectra, is
obtained. The calculations yield a nonmagnetic ground state that is
characterized by a noninteger filling of the plutonium 5f shell. The
noninteger filling as well as the satellites appearing in the 4f
core-level photoemission spectra originate in a sizable hybridization
of the 5f shell with the 2p states of oxygen.
\end{abstract}

\section{Introduction}

Plutonium dioxide is a correlations-driven
insulator~\cite{mcneilly1964,mccleskey2013} with a temperature
independent magnetic susceptibility~\cite{raphael1968}. It
crystallizes in the CaF$_2$ structure (space group Fm$\bar{3}$m), with
eight-coordinated plutonium, and four-coordinated oxygen atoms. Due to
the prominent role of electron correlations, the theoretical modeling
of the electronic structure of PuO${}_2$ presents numerous challenges.

The conventional Kohn--Sham density-functional theory in the
local-density (LDA) and generalized gradient approximations fails to
explain the insulating character of the oxide~\cite{wen2013}. The
band-gap problem was addressed a number of times using
orbital-dependent functionals such as the self-interaction corrected
LDA~\cite{petit2010}, LDA+U~\cite{suzuki2013}, or the hybrid
exchange-correlation functionals~\cite{wen2013,wen2012}. All of these
calculations lead to an insulator with large magnetic moments at the
plutonium atoms in the ground-state solution, in disagreement with
experiments. Such overestimated tendency to magnetism appears to be a
rather general shortcoming of static mean-field approximations that
build on a single determinant of Kohn--Sham orbitals.

The dynamical mean-field theory (DMFT) is able to describe correlated
nonmagnetic phases~\cite{georges1996}. This method, in combination
with the density-functional theory, was successfully applied to
PuO${}_2$ recently \cite{yin2011,shick2014}, and it indeed yields an
insulating electronic structure without local magnetic moments
\cite{shick2014}. In the present paper, we follow up on our previous
study \cite{shick2014} where we employed a crystal-field potential
deduced from experiments, and assumed a simplified spherically
symmetric hybridization of the plutonium 5f shell with the surrounding
states. Here we relax these simplifications and determine the
quantities entirely from first principles.

\section{Methods}



We start our investigation with an LDA calculation \cite{perdew1992}
of the electronic structure of PuO${}_2$ at the experimental lattice
constant 5.396 \AA~\cite{PearsonsHandbook} taking into account
scalar-relativistic effects as well as the spin-orbital coupling. To
this end, we employ the \emph{WIEN2k} package \cite{wien2k} with the
following parameters: the radii of the muffin-tin spheres are $R_{\rm
MT}({\rm Pu})=2.65\, a_{\rm B}$ for plutonium atoms and $R_{\rm
MT}({\rm O})=1.70\, a_{\rm B}$ for oxygen atoms, the basis-set cutoff
is defined with $R_{\rm MT}({\rm O})\times K_{\rm max}=8.5$, and the
Brillouin zone is sampled with 3375 k points (120 k points in the
irreducible wedge). The LDA bands of the plutonium 5f and oxygen 2p
origin are subsequently mapped onto a tight-binding model with the aid
of the \emph{Wannier90} code \cite{mostofi2008} in conjunction with
the \emph{Wien2wannier} interface~\cite{kunes2010}. This effective
model $\hat H_{\rm TB}$ then serves as a base for the LDA+DMFT
calculations.

The dynamical mean-field modeling of correlations among the 5f
electrons amounts to adding a local%
\footnote{It has non-vanishing
matrix elements only between the states of a given f shell.}%
selfenergy $\hat\Sigma(z)$ to the 5f shell of each Pu atom in the
tight-binding model $\hat H_{\rm TB}$. The selfenergy is taken from an
auxiliary impurity model that consists of one fully interacting f
shell (the impurity) embedded in a self-consistent non-interacting
medium $\bigl(\hat H_{\rm TB}+\hat\Sigma\bigr)$ \cite{georges1996}.

The auxiliary model without the f--f interactions can be written as
\begin{equation}
\label{eq:Himp0}
\hat H_{\rm imp}^{(0)}=
\sum_{ij}\bigl[\mathbb{H}_{\rm loc}\bigr]_{ij}\hat f_i^{\dagger}\hat
f_j
+\sum_{IJ}\bigl[\mathbb{H}_{\rm bath}\bigr]_{IJ}\hat b_I^{\dagger}\hat b_J
+\sum_{iJ}\Bigl(\bigl[\mathbb{V}\bigr]_{iJ}\hat f_i^{\dagger}\hat b_J
+\bigl[\mathbb{V}^{\dagger}\bigr]_{Ji}\hat b_J^{\dagger}\hat f_i
\Bigr)\,,
\end{equation}
where the lower-case indices label the f orbitals and run from 1 to
14, and the upper-case indices label the orbitals of the effective
medium that is usually referred to as bath. In the present
application, we truncate the bath to contain only 14 orbitals, in
which case the local hamiltonian~$\mathbb{H}_{\rm loc}$, the bath
hamiltonian $\mathbb{H}_{\rm bath}$ as well as the hybridization
$\mathbb{V}$ are all $14\times 14$ matrices. The truncation is
necessitated by the method we employ to solve the impurity model, the
exact diagonalization, which cannot handle much larger systems. In
insulating oxides like PuO${}_2$, such small bath is nevertheless well
justified on the physical grounds: the environment of the plutonium 5f
shell is dominated by the oxygen ligands and hence a small impurity
model analogous to a ligand-field model should accurately represent
the dynamics of the 5f shell and its surroundings.


The parameter matrices $\mathbb{H}_{\rm loc}$, $\mathbb{H}_{\rm bath}$
and $\mathbb{V}$ are determined by matching the large $z$ asymptotics
of the local Green's function of the impurity model $\hat H_{\rm
imp}^{(0)}$,
\begin{equation}
\label{eq:Gloc}
\mathbb{G}_{\rm loc}(z)=\Bigl[z \mathbb{I}-\mathbb{H}_{\rm loc}
 -\mathbb{V}\bigl(z \mathbb{I}-\mathbb{H}_{\rm bath}\bigr)^{-1}\mathbb{V}^{\dagger}
\Bigr]^{-1},
\quad\bigl[\mathbb{I}\bigr]_{ij}=\delta_{ij}\,,
\end{equation}
to the asymptotics of the local Green's function corresponding to the
effective medium $\bigl(\hat H_{\rm TB}+\hat\Sigma\bigr)$. The
procedure follows the steps outlined in \cite{kolorenc2012a} with a
notable difference that the local hamiltonian~$\mathbb{H}_{\rm loc}$ now
contains a strong spin-orbital coupling that does not commute with the
cubic hybridization function $\mathbb{V}\bigl(z
\mathbb{I}-\mathbb{H}_{\rm bath}\bigr)^{-1}\mathbb{V}^{\dagger}$.
Therefore, the problem cannot be simplified to diagonal matrices.

The complete interacting impurity model is defined as $\hat
H_{\rm imp}=\hat H_{\rm imp}^{(0)}+\hat U$ where $\hat U$ is the
Coulomb repulsion among the f orbitals,
\begin{equation}
\label{eq:vertex}
\hat U=
\frac12\sum_{m_1m_2m_3m_4\sigma\sigma'} U_{m_1m_2m_3m_4}
\hat f_{m_1\sigma}^{\dagger}\hat f_{m_2\sigma'}^{\dagger}
\hat f_{m_4\sigma'}\hat f_{m_3\sigma}
-U_{\rm H}\sum_{i}\hat f_i^{\dagger}\hat f_i\,.
\end{equation}
The matrix elements $U_{m_1m_2m_3m_4}$ are parametrized with the
Slater integrals $F_0=6.5$~eV, $F_2=8.1$~eV, $F_4=5.4$~eV and
$F_6=4.0$~eV. The average Coulomb repulsion $U=F_0$ has the same value
as in \cite{yin2011} to facilitate comparison of the calculated
spectra, the other integrals are chosen to obtain the average exchange
$J=0.7$~eV while keeping the Hartree--Fock ratios $F_4/F_2$ and
$F_6/F_2$ \cite{moore2009}. The second term in eq.~\eqref{eq:vertex}
is the double-counting correction that removes the Hartree-like
contribution already included in the LDA bandstructure $\hat H_{\rm
TB}$. We approximate $U_{\rm H}$ by the so-called fully localized
limit $U_{\rm H}=U(n_f-1/2)-J(n_f-1)/2$ \cite{czyzyk1994,solovyev1994}
where $n_f$ is the self-consistently determined number of 5f electrons.

The selfenergy $\hat\Sigma(z)$ of the impurity model $\hat H_{\rm
imp}$ is computed using an in-house exact-diagonalization code that
combines the implicitly restarted Lanczos method for calculation of
the bottom of the many-body spectrum \cite{arpack} with the band
Lanczos method for evaluation of the one-particle Green's function
\cite{meyer1989}. To lessen the computational demands, the Fock space
is reduced in a manner analogous to the method developed for Ce
compounds by Gunnarsson and Sch\"onhammer \cite{gunnarsson1983}. The
many-body basis $|f^n\text{\underline{$b$}}^m\rangle$, where $n$
indicates the number of electrons in the f~states and $m$ is the
number of holes in the bath states, is truncated at $m\leq N_{\rm
  h}$. We find that the quantities of interest are essentially
converged when the cutoff $N_{\rm h}$ is three or larger.

\subsection{Photoemission spectra}

The valence-band photoemission intensity can be estimated using the
Fermi's golden rule as
\begin{equation}
\label{eq:valence_spectrum}
w_{\rm v}(E) \sim \Im\Tr\sum_k\Bigl[
(E-\rmi \gamma_{\rm v})\hat I-\hat H_{\rm TB}(k)
-\hat\Sigma(E-\rmi \gamma_{\rm v})\Bigr]^{-1}\,,
\end{equation}
where the sum runs over the Brillouin zone, $\hat I$ is the identity
operator, and $\gamma_{\rm v}$ models the life-time broadening of the
valence states together with the experimental resolution. In addition,
the impurity model of the dynamical mean-field theory provides a means
to calculate the photoemission from core levels
\cite{kim2004,cornaglia2007,hariki2013}. The photoelectron intensity
can be approximated by~\cite{gunnarsson1983,degroot2008}
\begin{equation}
\label{eq:core_spectrum}
w_{\rm c}(E)\sim\Im\,
\langle\text{g.s.}|\biggl[
\bigl(E-\rmi\gamma_c-E_{\text{g.s.}}-\epsilon_{\rm c}\bigr)\hat I
+\hat H_{\rm imp}-U_{\rm
  cv}\sum_i\hat f^{\dagger}_i\hat f_i\biggr]^{-1}
|\text{g.s.}\rangle\,,
\end{equation}
where $|\text{g.s.}\rangle$ is the ground state of $\hat H_{\rm imp}$,
$E_{\text{g.s.}}$ is the corresponding ground-state energy, and
$\gamma_{\rm c}$ simulates the life-time broadening of the core
state. The term $-U_{\rm cv}\sum_i\hat f^{\dagger}_i\hat f_i$
represents the Coulomb attraction between the core hole and the 5f
electrons, and $\epsilon_{\rm c}$ is the energy of the core electron
that is removed in the final state. The many-body satellites appear in
the core spectra because the initial state of the photoemission
process $|\text{g.s.}\rangle$ is not an eigenstate of the final-state
hamiltonian 
$\bigl(\hat H_{\rm imp}
-U_{\rm cv}\sum_i\hat f^{\dagger}_i\hat f_i
-\epsilon_{\rm c}\hat I\bigr)$
due to non-commutativity of the number operator $\hat
f^{\dagger}_i\hat f_i$ with the hybridization terms~$\hat
f_i^{\dagger}\hat b_J$ and~$\hat b_J^{\dagger}\hat f_i$.

\section{Results}
\medskip

\subsection{Valence-band spectrum}

\begin{figure}
\begin{minipage}[t]{0.46\linewidth}
\includegraphics[width=\linewidth]{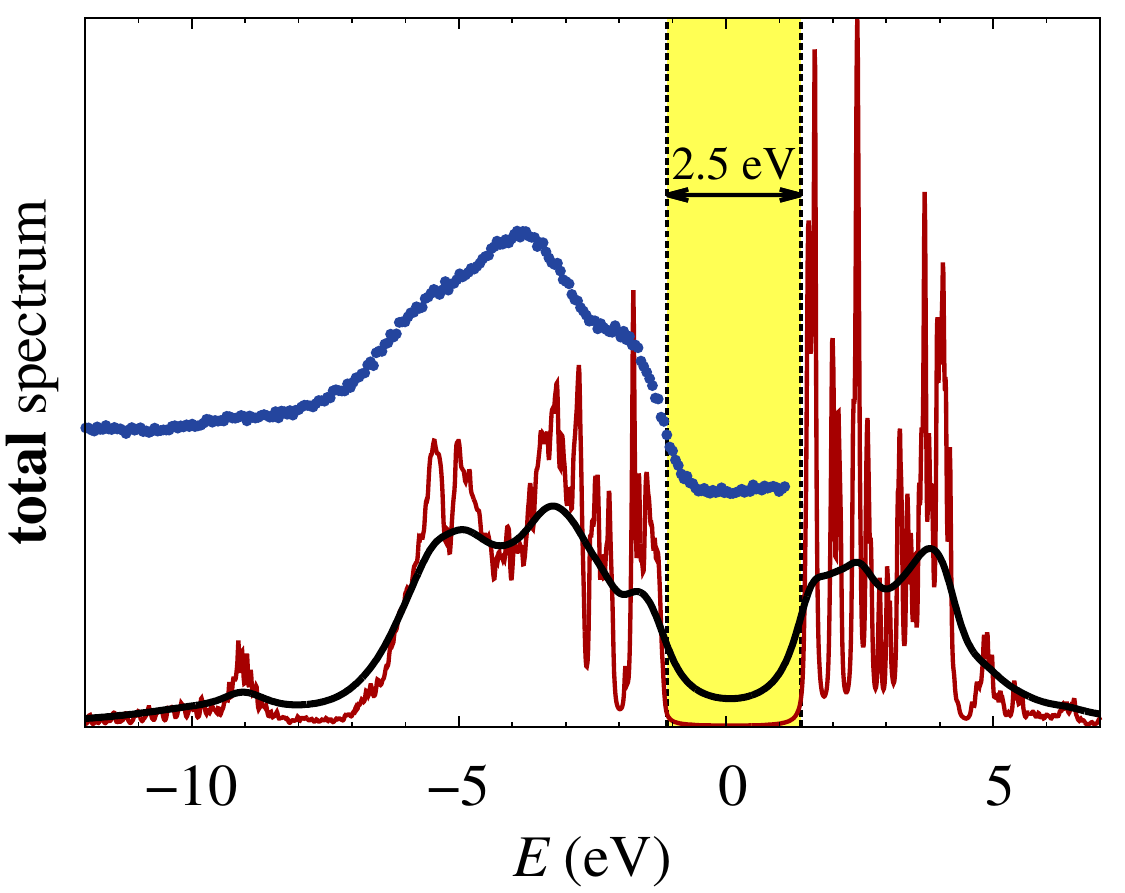}\vskip-.7em%
\caption{\label{fig:dos_tot}Valence spectrum $w_{\rm v}$
  from eq.~\eqref{eq:valence_spectrum} with broadening
  $\gamma_{\rm v}=0.02$ eV ({\color{dred}\full}) and $\gamma_{\rm
    v}=0.4$ eV ({\thickfull}). The experimental data
  ({\color{dblue}\dotted}) taken with incident photon 
    energy 40.8~eV (He~II line) \cite{gouder2013} are shown for comparison.}
\end{minipage}\hskip.08\linewidth%
\begin{minipage}[t]{0.46\linewidth}
\includegraphics[width=\linewidth]{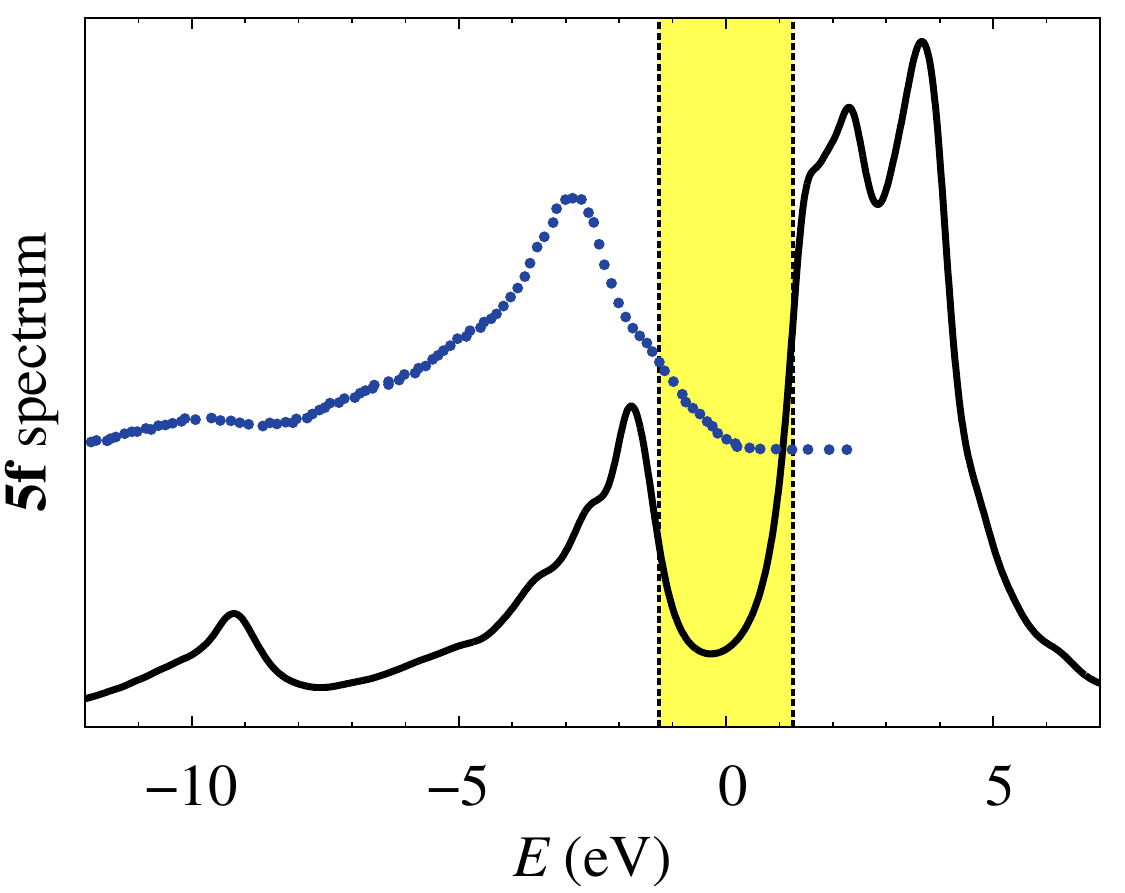}\vskip-.7em%
\caption{\label{fig:dos_f}5f component of the valence-band spectrum
  calculated with broadening $\gamma_{\rm v}=0.4$ eV ({\thickfull}) is
  compared to the experiment ({\color{dblue}\dotted}) performed at
  incident photon energy 1487 eV (aluminum K${}_{\alpha}$ line)
  \cite{teterin2013}.}
\end{minipage} 
\end{figure}

The theoretical spectrum of PuO${}_2$ is shown in
figure~\ref{fig:dos_tot} for two values of the life-time
broadening~$\gamma_{\rm v}$. The smaller value is used to visualize
the band gap that comes out as 2.5~eV. This gap agrees rather well
with 2.8~eV found in recent measurements of optical absorption
\cite{mccleskey2013}, which justifies our choice of Coulomb
$U=6.5$~eV.  The spectrum calculated with the larger broadening is
compared with the experimental photoemission measured with incident
photons of energy 40.8~eV (He~II line) \cite{gouder2013}. At this
energy, the photoionization cross section for plutonium 5f and oxygen
2p states is approximately the same and hence the photoelectron
intensity corresponds to the total spectrum. The spectrum has three
main features that are satisfactorily reproduced by the theory. The
calculations place these features at energies $-1.7$~eV, $-3.2$~eV and
$-5.0$~eV. An inspection of the orbitally resolved spectra indicates
that the shoulder at $-1.7$~eV is dominated by the 5f states whereas
the other two peaks are due to the 2p states. This assignment is
further corroborated by photoemission spectra measured at high photon
energy (1487~eV, aluminum K${}_{\alpha}$ line) \cite{teterin2013}
where only the 5f states contribute. The experiment also detects a
satellite of the 5f origin that our calculations put at approximately
$-9.0$~eV; the experimental spectrum appears to be shifted by about
1~eV to higher binding energies.

\subsection{4f spectrum}

\begin{figure}[b]
\includegraphics[width=0.46\linewidth]{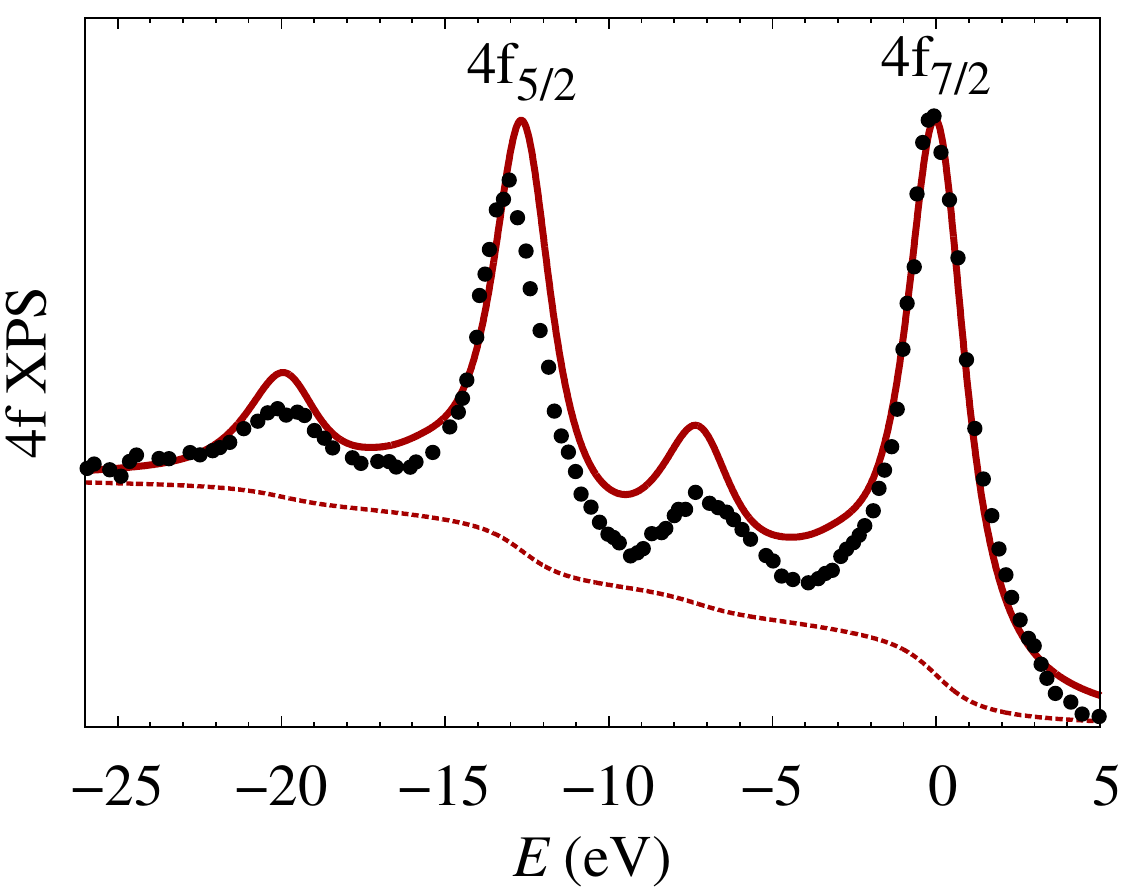}\hskip.08\linewidth%
\begin{minipage}[b]{0.46\linewidth}\caption{\label{fig:xps}Photoemission
    from the plutonium 4f levels calculated as a superposition of two spectra
    from eq.~\eqref{eq:core_spectrum} weighted with the statistical
    ratio $4/3$ ({\color{dred}\thickfull}). The 4f${}_{5/2}\,$--$\,$4f${}_{7/2}$
    splitting is taken from the LDA calculation (12.64 eV), the
    core-valence potential $U_{\rm 
      cv}$ is set to 6~eV, and the broadening $\gamma_{\rm c}$ is
    adjusted to match the width of the 4f$_{7/2}$ line in the
    experimental spectrum (\dotted) \cite{veal1977}. The calculated
    and experimental spectra are aligned at the 4f$_{7/2}$ line, and a
    background due to secondarily scattered electrons
    ({\color{dred}\shortdash}) is added to the theoretical curve as
    described in \cite{kotani1992}.}
\end{minipage}
\end{figure}

From the hybrid density-functional calculations \cite{wen2013,wen2012}
it was inferred that there is a large covalent mixing of the plutonium
5f states with the 2p states of oxygen. Evidence of such
mixing is found also in the core-level spectroscopy
\cite{teterin2013,veal1977}. Its theoretical interpretation
\cite{kotani1992} indicates that the filling of the plutonium 5f shell
noticeably departs from the nominal integer occupation $n_f=4$
associated with the Pu${}^{4+}$ ion.


Our calculations yield $n_f\approx 4.4$ which is in line with the
earlier studies although slightly smaller than $n_f=4.65$ deduced in
\cite{kotani1992}. The comparison of the 4f photoemission spectra
obtained experimentally \cite{veal1977} and calculated from
eq.~\eqref{eq:core_spectrum}, figure~\ref{fig:xps}, shows that our
modeling of the local electronic structure at the plutonium atom
is sufficiently accurate. In particular, the satellites that
originate from the hybridization with the oxygen ligand states are
well reproduced without any fitting of the hybridization strength --
it comes directly from the LDA+DMFT calculations.

\subsection{Crystal-field states}

Finally, we discuss the role of the cubic environment around the
plutonium atoms. Its most important implication is the absence of
magnetism \cite{raphael1968} -- the crystal-field potential and the
hybridization split the 5f states such that the ground state is
non-degenerate. The first excited state is a triplet and it was found
at 124~meV above the ground state in inelastic neutron scattering
\cite{kern1999}. The first excitation in our impurity model $\hat
H_{\rm imp}$ turns out to be at 125~meV which is an excellent
agreement but more compounds will have to be tested to check whether
such accuracy is systematic or somewhat fortuitous.
 
To analyze the splitting in more detail, we decompose the local
hamiltonian $\mathbb{H}_{\rm loc}$ into a sum of the spin-orbital term
$\xi(\mathbf{l}\cdot\mathbf{s})$ and the cubic crystal-field potential
\begin{equation}
V_{\rm CF}= \frac{16\sqrt{\pi}}{3}  V_{4}
\biggl[Y_{40}(\theta,\varphi) +
\sqrt{\frac{10}{7}} \Re Y_{44}(\theta,\varphi)\biggr]
 +   32 \sqrt{\frac{\pi}{13}} V_{6} \Bigl[Y_{60}(\theta,\varphi) - \sqrt{14}
 \Re Y_{64}(\theta,\varphi)\Bigr]\,,
\label{eq:CF}
\end{equation}
where $Y_{lm}(\theta,\varphi)$ are spherical harmonics. The parameters
$\xi$, $V_4$ and $V_6$ are compared in table~\ref{tab:cf} with values
estimated by other methods: the inelastic neutron scattering
experiments \cite{kern1999} and the LDA+U method combined with the
crystal-field model \cite{zhou2012}. Apparently, our $V_4$ and $V_6$
are not directly comparable to the values from the literature. The
reason is that we employ an impurity model that explicitly includes
the hybridization with ligands whereas the other studies work with a
simpler crystal-field model where the hybridization term is absent and
its effect is folded into the crystal-field potential. By switching
off the hybridization while keeping our small values of $V_4$ and
$V_6$ we find that the hybridization is responsible for approximately
half of the splitting due to the cubic environment.

\begin{table}
\lineup
\caption{\label{tab:cf}%
Parameters of the local hamiltonian $\mathbb{H}_{\rm loc}$, and the
first three excitation energies (the number in brackets indicates the
degeneracy of the excited state). Our calculations are compared with
results of two other methods. The states $\Gamma_5(3)$ and $\Gamma_3(2)$
are interchanged in our calculations compared to LDA+U
\cite{zhou2012}. All energies are listed in meV.}
\begin{center}
\begin{tabular}{l@{\qquad}cc@{\hskip1.4em}c@{\hskip2.8em}ccc}
\br
 & $\xi$ & $V_4$ & $V_6$ & $\Gamma_4(3)$ & $\Gamma_5(3)$ & $\Gamma_3(2)$\\
\mr
inelastic neutrons \cite{kern1999} & 300 & $-151$ & 31 & 124 \\
LDA+U \cite{zhou2012}              & 304 &\0$-99$ & 17 &\097 & 204 & 195 \\
present study                      & 322 &\0$-39$ & 12 & 125 & 226 & 319 \\
\br
\end{tabular}
\end{center}
\end{table}

\section{Conclusions}

We have demonstrated that a variant of the LDA+DMFT method where the
selfenergy is obtained by exact diagonalization of a finite impurity
model similar to the multiplet ligand-field model provides an accurate
and comprehensive description of the electronic structure of the
plutonium dioxide. A nonmagnetic insulator is obtained, and the main
features of the valence-band as well as core-level photoemission
spectra are well reproduced. The method allows us to quantitatively
analyze the effects of hybridization between the plutonium 5f and
oxygen 2p states from first principles.

\ack

We acknowledge financial support from the Czech Republic Grant GACR
P204/10/0330. A.~L.~K. acknowledges financial support from the
International Visegrad Fund within the research project
No. 51301118. Access to computing and storage facilities owned by
parties and projects contributing to the National Grid Infrastructure
MetaCentrum, provided under the programme ``Projects of Large
Infrastructure for Research, Development, and Innovations''
(LM2010005), is greatly appreciated.


\raggedright
\bibliography{dft,dmft,xpsDMFT,lanczos,aim,codes,AnO,pu}

\end{document}